# Snapping for high-speed and high-efficient, butterfly swimming-like soft flapping-wing robot

Yinding Chi[1], Yaoye Hong[1], Yao Zhao[1], Yanbin Li[1], Jie Yin[1*]

[1]Department of Mechanical and Aerospace Engineering, North Carolina State University, Raleigh, NC 27695 USA

*Corresponding author: jyin8@ncsu.edu

**Abstract:** Natural selection has tuned many flying and swimming animals across different species to share the same narrow design space for optimal high-efficient and energy-saving locomotion, e.g., their dimensionless Strouhal numbers St that relate flapping frequency and amplitude and forward speed fall within the range of 0.2 < St < 0.4 for peak propulsive efficiency. It is rather challenging to achieve both fast and high-efficient soft-bodied swimming robots with high performances that are comparable to marine animals, due to the observed narrow optimal design space in nature and the compliance of soft body. Here, bioinspired by the wing or fin flapping motion in flying and swimming animals, we report leveraging the generic principle of snapping instabilities in the bistable and multistable flexible pre-curved wings for high-performance, butterfly swimming-like, soft-bodied flapping-wing robots. The soft swimming robot is lightweight (2.8 grams) and demonstrates a record-high speed of 3.74 body length/s (4.8 times faster than the reported fastest soft swimmer), high-efficient (0.2 < St = 0.25 < 0.4), low energy consumption cost, and high maneuverability (a high turning speed of 157°/s). Its high performances largely outperform the state-of-the-art soft swimming robots and are even comparable to its biological counterparts.

## INTRODUCTION

Flapping motion is a fast yet energy-efficient locomotion mode in flyers and swimmers such as birds, insects, and marine animals (*1, 2*). They leverage bending and/or rotating flexible wings, fins, body, or tails for passively increasing propulsion efficiency to save energy (*2*). Among them, many are observed to cruise in a narrow range of dimensionless Strouhal numbers St, defined as St = $fA/U$ ($f$ and $A$ are flapping frequency and amplitude and $U$ is forward velocity), i.e., 0.2 < St < 0.4, for high power efficiency (*3, 4*). Bioinspired aqueous soft robots utilize similar flapping or oscillation motion driven by various fluidic, electrical, and photo-active soft actuators for propulsion (*5-8*). However, their performances are far from competing with marine animals from the perspectives of both speed (< 1 body length/ second (BL/s) (*5*) in soft robots *vs.* 2-24 BL/s (*9*) in marine animals) and propulsion efficiency (St > 1 or St < 0.1 in soft robots *vs.* 0.2 < St < 0.4 in marine animals) (*6-8, 10, 11*). It remains a grand challenge to achieve both fast and high-efficient aqueous soft robots with high performance comparable to their biological counterparts, due to the compliance of their soft body (*12*) and the naturally selected narrow design space (St) for high propulsive efficiency.

To address the challenge, here, we exploit leveraging snapping of bistable flexible wings for amplified aqueous performances in a soft-bodied flapping-wing swimming robot (**Fig. 1A**). Snapping is a fast motion often observed in nature (e.g., fast closure of Venus flytraps (*13*)) and daily life (e.g., popper jumping toys and hair clippers). Harnessing snapping for high-performance soft robots has recently attracted growing interest in addressing compliance-related issues (*5, 14-20*). However, soft swimming robots with comparable high performances to their biological counterparts have yet to be realized (*5-8, 10, 11, 21-25*). We present a generic design of a bistable and multistable soft-bodied flapping-wing swimmer composed of soft bending actuators as soft body and a pair of pre-curved flexible wings (**Fig. 1B** and **fig. S2**). Inspired by the design of hair clippers, bonding two parallel wing frame ribbons at the tip forms a pair of bistable pre-curved flexible wings (**Fig. 1B**), resulting from lateral torsion and compression induced out-of-plane buckling. The actuated small flexion of the soft body can drive the passive snapping of the wings for largely amplified flapping and rotating performances. We explore the underlying generic design principle and their dynamic performances. Leveraging the knowledge, we exploit harnessing bistability and multistability for high-speed, high-energy-efficient, and maneuverable soft swimmers.

## RESULTS

### Figure-of-eight-like flapping wing and amplified wing performances

**Fig. 1A** shows the passive snapping and flapping of the bistable flexible wings (wingspan length $S = 150$ mm) from both side (*XZ* plane) and front views (*YZ* plane) during one cycle of downstroke and upstroke wing motions (Supplementary **Video S1**). One end of the soft pneumatic bending actuator-based body is clamped (actuation pressure is 55 kPa and actuation frequency is 0.714 Hz, Table S1). The bending and slightly elongation deformation in the soft body drive the simultaneous clockwise rotation and flapping of both wings (**Fig. 1A**). This is in sharp contrast to conventional soft flapping or oscillation actuators that only undergo bending rather than combined bending and rotation here (*5-8, 10, 24*). Consequently, the trajectory of the wing tip shows an interesting figure-of-eight-like profile (**Fig. 1C**), similar to that of the flapping wings observed in a hovering hummingbird and bumblebee for augmented thrust force (*26, 27*) (**Fig. 1D-E**). Differently, the cross shape is steep here. It represents the fast snap-through from state ii to iii (within 40 ms) during downstroke and the snap-back from state vi to vii (within 36 ms) during upstroke (**Fig. 1A**). During snapping, both the wing tip's speed and acceleration rate increase dramatically to an instantaneous peak high speed of $v_{max-x} \approx v_{max-z} \sim 6.6$ m/s and rate of acceleration $a_{max-x} \approx a_{max-z} \approx 1.49 \times 10^3$ m/s$^2$ **fig. S3**), which is orders of magnitude higher than the gravitational acceleration. To realize such a sophisticated figure-of-eight-like trajectory in flapping robots, it often requires a complex multiple bar-linkage transmission system between the actuator and the wing (*28, 29*). In contrast, our flapping and rotating motion is transmission-free that largely simplifies the design.

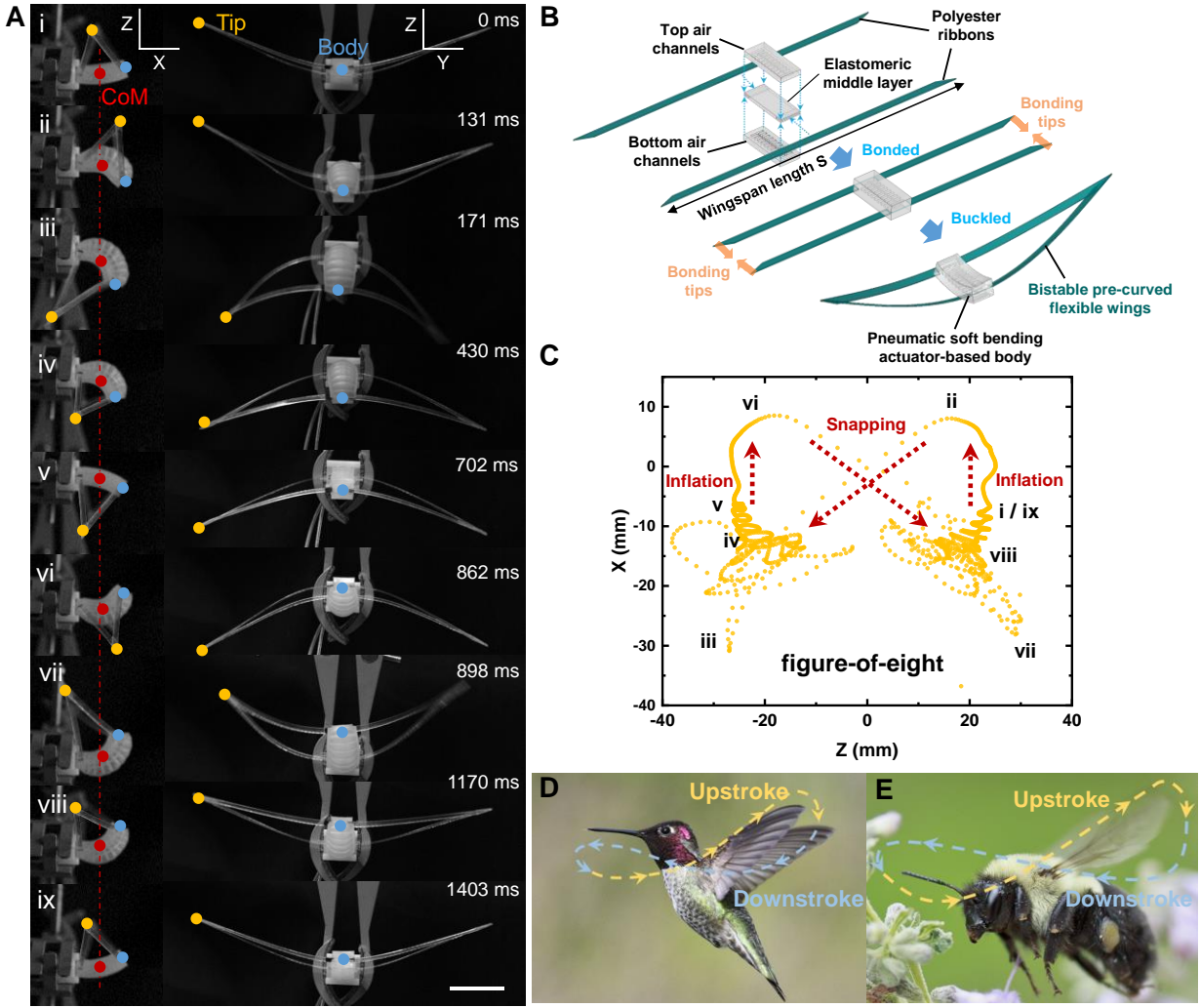

**Fig.1 Snapping-induced amplified flapping and rotating motions in the bistable pre-curved soft flapping actuator (wingspan length $S$ = 150 mm).** (**A**) The time-lapse images of pneumatic actuated motions during a representative cycle of downstroke (snap through, state ii-iii in **A**) and upstroke (snap back, state vi-vii in **A**) wing flapping in both side view (left, *XZ* plane) and front view (right, *YZ* plane) captured by a high-speed camera. The orange and cyan dots denote the wing tip and soft body as motion trackers. The red dot denotes the center of mass (CoM) of the soft body. The scale bar: 20 mm. (**B**) Schematic illustration of fabrication process of the bistable pre-curved soft bending actuator. (**C**-**E**) Trajectory of the wing tip in *XZ* plane follows a figure-of-eight-like loop (**C**), similar to that of the hovering images of a hummingbird (**D**) and bumblebee (**E**) at their wing tips.

Snapping can largely amplify both rotating and flapping motion of the wings through sudden release of the stored strain energy, whereas it only requires relatively small bending deformation in the soft body with low actuation energy. To better understand the connection between the pneumatic actuated deformation in the soft body and the induced amplified flapping and rotating behavior in the wing (**Fig. 2A**), we track the bending angle $\varphi_{body}$ and deflection $d_{body}$ of the soft body as a function of the actuation pressure as shown in **Fig. 2B** and **Fig. 2C**, respectively, as well as the corresponding changes in the rotation angle $\varphi_{wing}$ (**Fig. 2D**), flapping angle $\theta_{wing}$ (**Fig. 2E**), and deflection $d_{wing}$ of the wing (**Fig. 1C**).

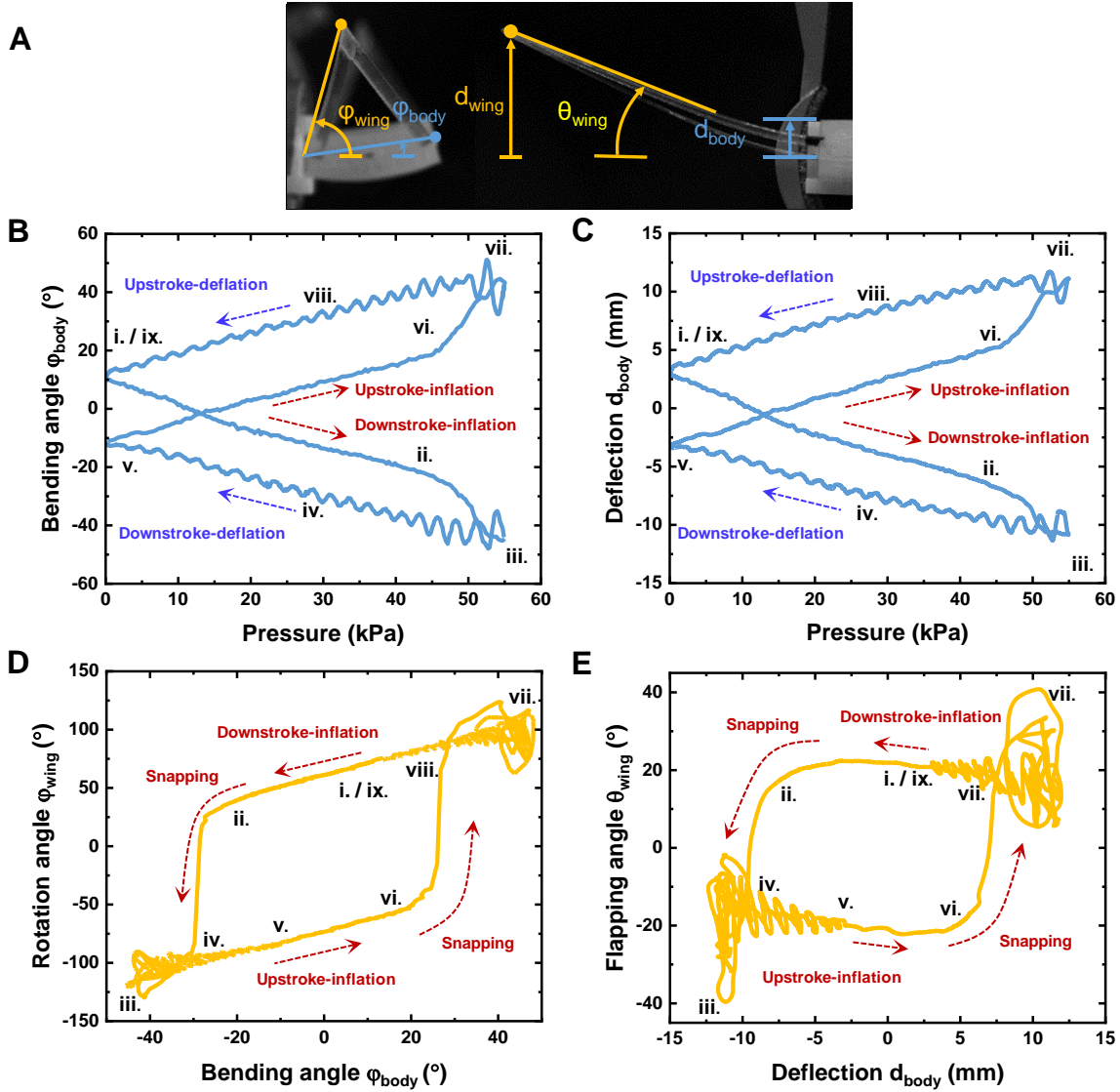

**Fig. 2. Quantitative relationship between the actuated deformation in the soft body and the passive deformation in the bistable flexible wings during a single flapping cycle of downstroke and upstroke ($S$ = 150 mm).** (**A**) Definitions of soft body bending angle $\varphi_{body}$ and wing rotating angle $\varphi_{wing}$ (left), wing deflection $d_{wing}$ and flapping angle $\theta_{wing}$ and soft body deflection $d_{body}$ (right). (**B-E**) The bending, rotating, and deflection performances of the flapping actuator. i-ix represent the nine representative dynamic deformed states shown in **Fig. 1A**. (**B-C**) Soft body bending angle $\varphi_{body}$ and deflection $d_{body}$ as a function of pneumatic pressure $p$. (**D**) Wing rotation angle $\varphi_{wing}$ as a function of $\varphi_{body}$. (**E**) Wing flapping angle $\theta_{wing}$ as a function of $d_{body}$.

Upon inflating the top chamber, i.e., downstroke from state i to ii prior to snapping, the soft body bends downward with its bending angle $\varphi_{body}$ decreasing approximately linearly from 10° to ~ -20° with the pressure, i.e., $\Delta^{i-ii}\varphi_{body}$ ~ 30° (**Fig. 2B**). Equivalently, its deflection $d_{body}$ decreases linearly from 2.5 mm to ~ -7.2 mm, i.e., $\Delta^{i-ii}d_{body}$ ~ 9.7 mm (**Fig. 2C**). Correspondingly, for the wings, its rotation angle $\varphi_{wing}$ decreases linearly from 75° to ~ 40°, i.e., $\Delta^{i-ii}\varphi_{wing}$ ~ 35° (**Fig. 2D**), whereas its flapping angle $\theta_{wing}$ remains almost unchanged as ~ 20°, i.e., $\Delta^{i-ii}\theta_{wing}$ ~ 1° (**Fig. 2E**), and its deflection decreases slightly with $\Delta^{i-ii}d_{wing}$ ~ 9.7 mm.

During the snap-through from state ii to iii, the small variation in the body deflection $\Delta^{ii-iii}d_{body}$ (~ 5.5 mm) triggers over 7 times larger wing deflection $\Delta^{ii-iii}d_{wing}$ (~ 38.8 mm = 0.26 $S$) (**Fig. 1C**). Bending of the soft body $\Delta^{ii-iii}\varphi_{body}$ (~ 22°) also induces a large rotation angle change $\Delta^{ii-iii}\varphi_{wing}$ over 170° (**Fig. 2D**) and a large flapping angle change $\Delta^{ii-iii}\theta_{wing}$ over 60° (**Fig. 2E**), which are over 4.8 times $\Delta^{i-ii}\varphi_{wing} \approx \Delta^{i-ii}\varphi_{body}$ due to the preserved soft body bending-wing rotation transmission and 60 times $\Delta^{i-ii}\theta_{wing}$ from state i to ii prior to snapping, respectively. Similar snapping-induced large amplified effects are also observed during the upstroke of the wing from state vi to vii in **Fig. 1C** and **Fig. 2B-E**.

**Tunable pre-curved shapes of the flexible bistable wings through wingspan length $S$**

As shown in the analytical model, experiments, and finite element analysis (FEA) simulation (Materials and Methods), the pre-curved wing shape and its flexibility can be tuned by the bending stiffness of the wing frame ribbon, the wingspan length $S$ (**Fig. 3A**), and the soft body length $L$ (**fig. S4**). The shape of the pre-curved wing can be characterized by two bending curvatures due to the tip bonding induced distortion, i.e., $\kappa_{XY}$ (or bending angle $\alpha$, inset of **Fig. 3B**) in the $XY$ plane and $\kappa_{YZ}$ (or bending angle $\gamma$, inset of **Fig. 3C**) in the $YZ$ plane. **Fig. 3B-3C** show that both bending curvatures decrease monotonically with the increase of $S$. As expected, the shorter the wingspan length, the higher pre-stress and larger curvature it generates, and the larger bending stiffness it possesses (**fig. S5A**). $S$ shows a more prominent effect on $\kappa_{XY}$ than $\kappa_{YZ}$. The corresponding FEA simulation results and analytical model agree well with the experiments (**Fig. 3A-3C**). **Fig. 3D** shows the theoretically predicted strain energy density $u$ as a function of $S$. It shows that as $S$ increases from 120 mm to 180 mm, $u$ decreases nonlinearly by over a half from about 1.09 to 0.48 mJ/mm, which is consistent with the corresponding FEA simulation results. The higher strain energy density at smaller $S$ also indicates the higher energy barrier, as evidenced by the enclosed larger area of the force-deflection curves of the bistable actuators in **fig. S5A**. Similarly, increasing the body length $L$, i.e., the parallel distance between the two ribbon frames prior to bonding, shows the similar effect as reducing $S$ on the wing shape change. Thus, geometrically, the larger the aspect ratio of $L/S$, the lager curvatures and pre-stored strain energy it will generate.

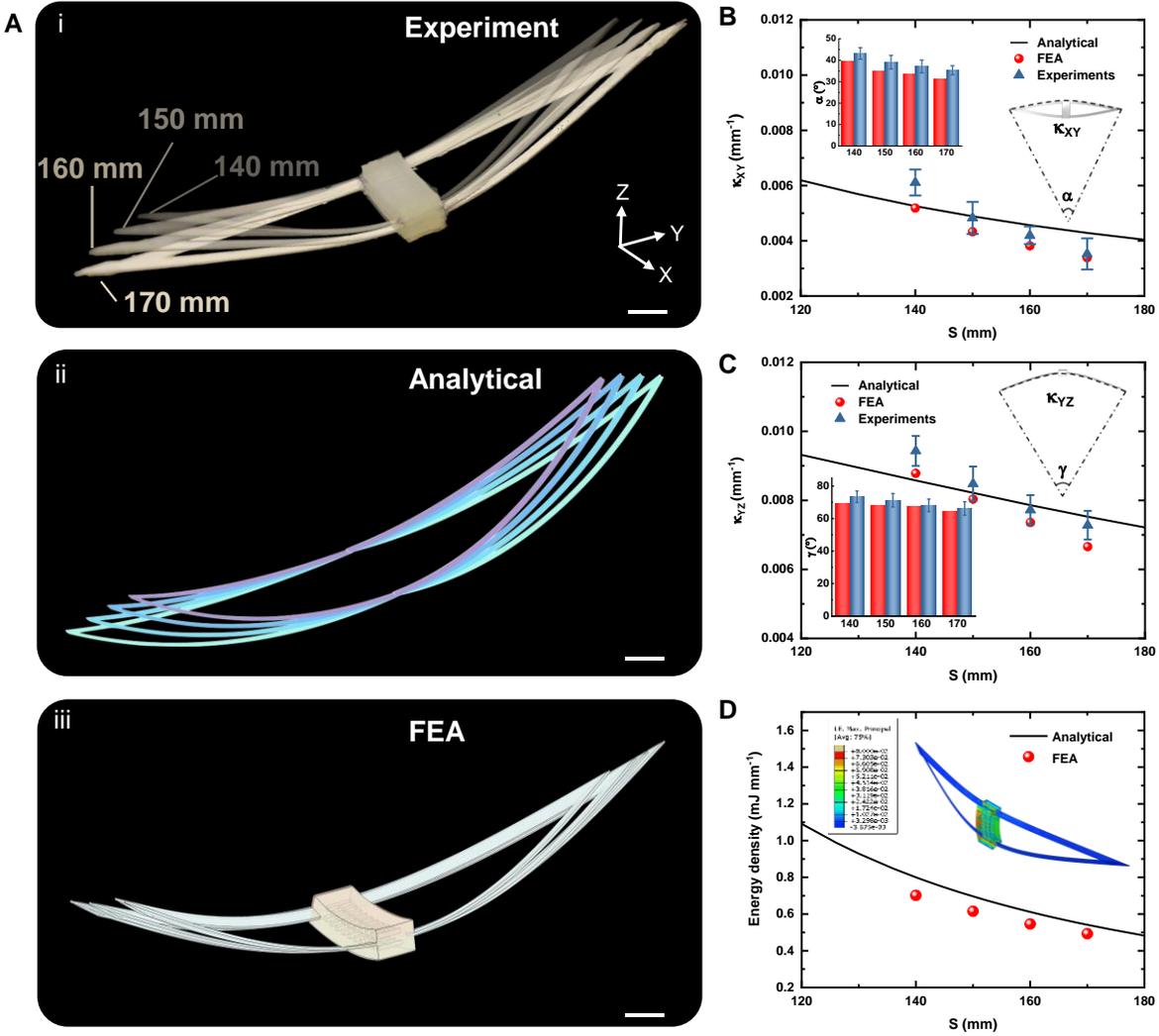

**Fig. 3. Tunable pre-curved shapes in the bistable soft flapping actuators with different wingspan length $S$.** (**A**) The overlapping images of the pre-curved bistable wings with different wingspan length $S$ ranging from 140 mm to 170 mm obtained from experiments (**i**), FEA simulation (**ii**), and theoretical prediction (**iii**). The scale bar: 10 mm. (**B - C**) The relationships between the curvatures and bending angles of the bistable flexible ribbons, $\kappa_{XY}$ and $\alpha$ in $XY$ plane (**B**) and $\kappa_{YZ}$ and $\gamma$ in $YZ$ plane (**C**), and their wingspan length $S$ from experiments, theory, and FEA simulation. The right insets show the schematics of the bending curvature and related bending angle. The left inset shows the measured bending angle change with $S$. (**D**) Comparison of elastic energy density of the bistable wings versus $S$ between theory and FEA simulation. The inset shows the simulated maximum principal strain contour for the bistable wing with $S = 150$ mm.

## Amplified dynamic actuated performances of bistable soft flapping actuators

**Figure 4A** and **Fig. 4B** show the front-view ($YZ$ plane) and side-view ($XZ$ plane) trajectories of the wing tip during one cycle of downstroke and upstroke motion with $S$ varying from 140 mm to 170 mm under the same actuation pressure (55 kPa) and frequency (0.714 Hz). The front view shows that the bending motions of all the wing tips follow a similar symmetric arc-shaped path (**Fig. 4A**). The bistable wings exhibit a similar large range of flapping angle $\Delta\theta_{wing} \sim 84°$, which is over 16 times larger than that of their monostable counterpart ($\Delta\theta_{wing} \sim 5.6°$) with flattened, stress-free wings (black curve in **Fig. 4A**), as well as over 8 times larger than that of the reported monostable soft electronic fish ($\Delta\theta_{wing} \sim 10°$) (*30*). Side-view trajectories show that all the bistable flapping wing tips follow a similar large figure-of-eight-like profile. In contrast, their monostable counterpart shows a small-angled segmental arch shape (black curve in **Fig. 4B**). The observed oscillated trajectory paths shown on the bottom are due to the vibration and damping after snapping (**Fig. 4B**).

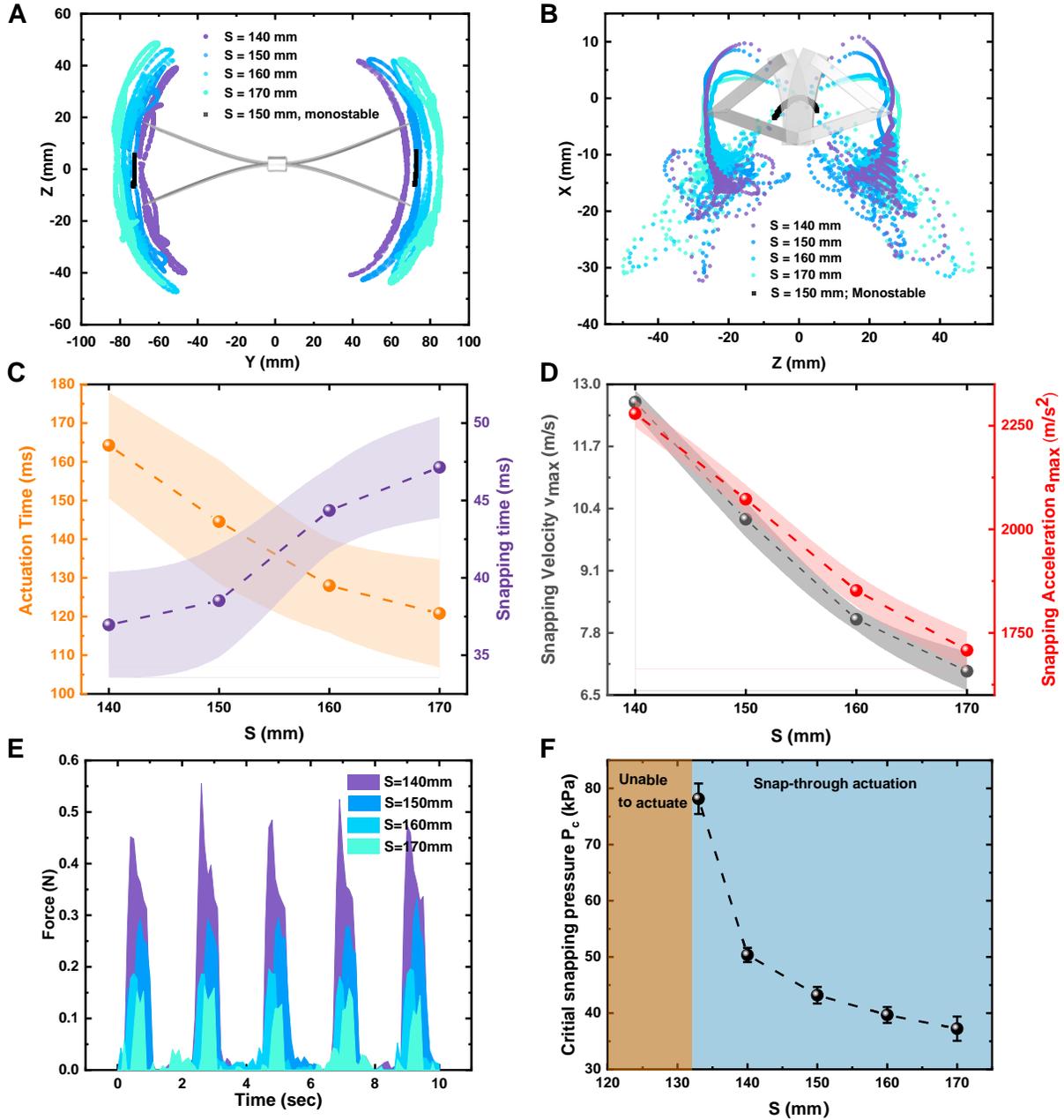

**Fig. 4. Comparison of the amplified dynamic performances in the bistable soft flapping actuators with different wingspan length *S*. (A -B)** Trajectories of the wing tips in *YZ* plane (front view of two wing tips, **A**) and *XZ* plane (side view of one wing tip, **B**) during one downstroke and upstroke flapping

cycle in both bistable and monostable (denoted by black curves) soft flapping actuators. **(C)** The actuation time and snapping time *vs*. *S*. All actuators are pressurized at 55 kPa with the same flow rate of 1.2 liters/min. **(D)** The maximum velocity and acceleration *vs*. *S* at the onset of snapping. **(E)** The dynamic block force changes with time during actuation. **(F)** The critical snapping pressure to trigger the onset of snapping *vs*. *S*. The gold region denotes the failure in activating the snapping ($S < 132.5$ mm), beyond that, the blue region denotes the activation zone.

The wingspan length *S* shows a profound effect on the dynamic flapping and actuation performances, including the stroke time and snapping time (**Fig. 4C**), wing tip's snapping velocity and acceleration rate (**Fig. 4D**), dynamic block force (**Fig. 4E**), and the critical actuation pressure $P_c$ for triggering snap-through instabilities (**Fig. 4F**). Generally, the shorter the *S* is, the longer actuation time and the faster snapping duration $t_{snap}$ it takes, and the higher snapping speed, snapping acceleration rate, and dynamic block force it generates, as well as the higher $P_c$ it requires to induce snapping as described below. **Fig. 4C** shows that as *S* decreases from 170 mm to 140 mm, it takes a shorter snap-through duration $t_{snap}$ that decreases from ~ 47 ms to ~ 37 ms, due to the faster energy release of the higher strain energy stored in the shorter *S*. Meanwhile, both snapping velocity $v_{max}$ and acceleration rate $a_{max}$ increase from ~ 7.0 m/s to ~ 12.6 m/s and from ~ 2,106 m/s$^2$ to ~ 2,764 m/s$^2$, respectively. Correspondingly, the dynamic block force is largely enhanced by over 3 folds that increases from ~ 0.15 N at $S = 170$ mm to ~ 0.52 N at $S = 140$ mm, as in **Fig. 4E**. Meanwhile, the achieved amplified force at the shorter *S* also requires a higher critical $P_c$ to overcome the higher energy barrier as shown in **Fig. 4F**. As *S* reduces to below 140 mm, $P_c$ increases steeply from ~ 50 kPa at $S = 140$ mm to ~ 78 kPa at $S = 132.5$ mm. This length also corresponds to the critical wingspan length for the current design, below which the pneumatic soft body fails to activate the snapping of the bistable wings even when it is overinflated (over 120 kPa) due to its dramatically increased energy barrier (**fig. S6**).

**Butterfly stroke-like, high-speed soft flapping-wing swimmer**

Next, we explore its potential applications in fast-speed and high-efficient swimming soft robots. As schematically illustrated in **Fig. 5A**, the bistable soft flapping-wing swimmer is constructed from simply covering the wings of the bistable soft flapping actuator in **Fig. 1A** with a flexible membrane, alongside attaching two flexible film-based extended fins to the trailing edges of both wings for amplifying the propulsion (see Materials and Methods for details). **Fig. 5B** and **Supplementary Video S4** show the side view of its swimming gaits during one cycle of downstroke and upstroke wing motion ($S = 150$ mm, actuation pressure 55 kPa and frequency 0.625 Hz). Similar to the wing flapping in air in **Fig. 1A**, the swimmer flaps and rotates its wings fast during snapping (e.g., a snapping period of 46 ms from $t = 0.225$ s to 0.271 s in **Fig. 5B**). It creates vortices behind (**Supplementary Video S5**) and pushes water backward to generate the thrust force and propel it forward. The body also shows an angled posture with its head diving deeply into water, as in **Fig. 5B** at $t = 0.817$ s during downstroke and $t = 1.073$ s during the upstroke.

We find that its swimming postures show similar characteristics to the most challenging butterfly swimming stroke in humans (*31*) (insets of **Fig. 5A** and **Fig. 5B**) in terms of both the body and arm motions. During the butterfly stroke, the human body undergoes undulating motion with the head leading the movement of the stroke. The wing-like arms provide the majority of the propulsion for the stroke by simultaneously stretching out, pulling, and sweeping both arms for propulsion (*32, 33*). Similarly, our soft flapping-wing swimmer bends its soft body to generate a wave-like undulation with its head moving up and down during the stroke, as shown in the undulating trajectory of the tracked body CoM (**Fig. 5C**), coordinating with the simultaneously fast strokes of flapping and sweeping both flexible wings via rotation for propulsion. Differently, during the recovery phase of the butterfly swimming stroke, the upstroke wing action in our swimmer can also provide the large thrust force to swim forward. Such a burst swimming mode is further verified by monitoring the instantaneous velocity changes during swimming, as in **Fig. 5D**. It shows a peak instantaneous high swimming speed of ~ 0.45 m/s during the snapping of downstroke.

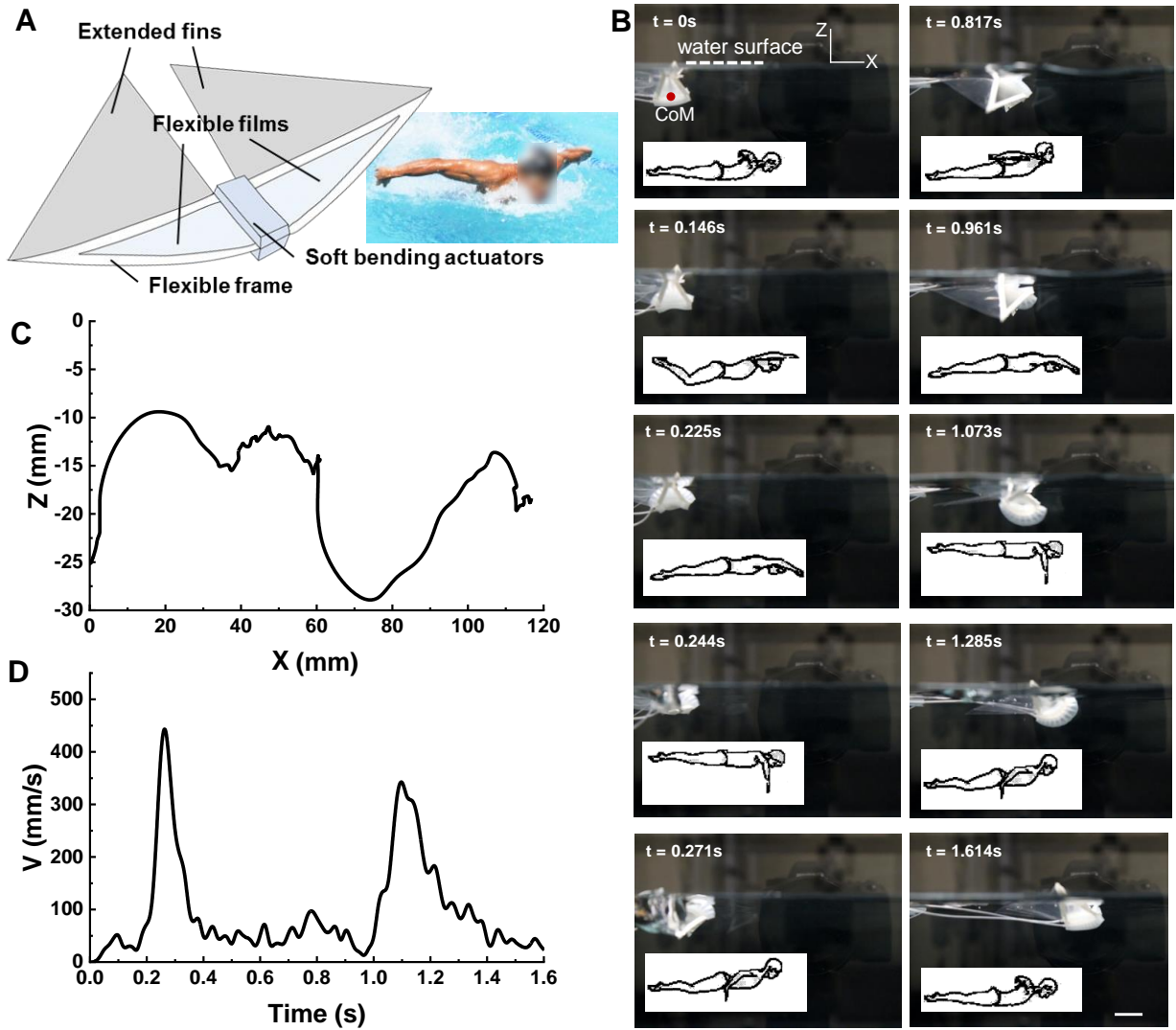

**Fig. 5. Butterfly stroke-like bistable flapping-wing soft swimming robot.** (**A**) Schematic of the proposed bistable soft swimming robot by attaching the bistable flapping soft actuator with flexible films and extended fins.The inset shows the butterfly swimming stroke. (**B**) The time-lapse side-view images of the representative swimming gaits in the bistable swimmer with wingspan length $S$ = 150 mm during one cycle of down- and up-stroke under actuation pressure of 55 kPa and frequency of 0.625 Hz. The insets shows the schematic human's swimming postures of the butterfly swimming stroke (*31*). The scale bar:

15 mm. At $t = 0$ s, the floating swimmer is fully immersed in water except the wing tips with its wings flapped upward and body angled slightly. Upon inflating, the body starts to bend downward and barely moves forward until beyond the onset of snapping at $t = 0.225$ s, two wings quickly and simultaneously stroke downward and rotate within 46 ms (from $t = 0.225$ s to $t = 0.271$ s). Consequently, it lifts the body close to the water surface ($t = 0.271$ s) and propels forward. The body shows an angled posture with its head diving into the water ($t = 0.817$ s). Upon inflating to bend the body upward to snapping both wings back, the fast wing upstroke and rotation make the body steeply dive into the water ($t = 1.073$ s), propelling it forward with its head floating up ($t = 1.285$ s). The swimmer recovers to its initial posture at $t = 1.614$ s. **(C)** The tracked center of mass (CoM) motion of its soft body. **(D)** The velocity profile of the bistable butterfly-stroke-like soft swimmer.

**Fast and lower energy-cost swimming performances**

Next, we further explore the effects of wingspan length $S$ and actuation frequency $f$ on its swimming performance (**Fig. 6A-6B** and **Supplementary Video S6**) under the same pressure of 55 kPa. **Fig. 6A** and **Supplementary Video S6** show the comparison of their swimming performances for the bistable flapping-wing swimming robots with $S$ varying from 140 mm to 170 mm under the same $f = 0.67$ Hz. It shows that the swimmer with an intermediate $S = 150$ mm swims the fastest, as also observed under different actuation frequencies (**Fig. 6B**), where their swimming speeds increase monotonically with $f$. The highest speed observed in the case of an intermediate $S = 150$ mm can be qualitatively explained as follows. For the swimmers with a shorter $S$, their higher pre-stored strain energy and the resulting higher block force from the wings are compromised by their relatively smaller interaction surface area with fluids (**fig. S7**), which lowers its thrust force and swimming speed (*34*). Similarly, for the swimmers with a longer $S$, they possess larger interaction surface areas whereas generate a smaller flapping force for propulsion due to the lower pre-stored energy. We note that unlike the fast bistable swimmers, their monostable counterpart ($S = 150$ mm) can barely move due to its non-amplified, much smaller flapping and rotation angles (**Fig. 4A-4B**) (**Supplementary Video S7**). Such intermittent swimming or burst-and-coast swimming style can generate higher thrust force for fast swimming speed than the undulating swimming mode at low-frequency domain (*35*). We find that the bistable soft swimmer with $S = 150$ mm actuated at 55 kPa and 1 Hz achieves the maximum fast speed of 85.27 mm/s, which corresponds to 3.74 BL/s (orange star in **Fig. 6B** and **Supplementary Video S8**).

Similar to the effect of $S$ on the swimming speed, the fastest soft swimmer with an intermediate $S = 150$ mm also shows the lowest cost of the transport (CoT) that quantifies the energy efficiency with lower CoT indicating lower energy consumption (**Fig. 6C**). CoT = $E / (m \times g \times d)$ is to quantify and evaluate energy consumption of transporting a target object with a certain distance, where $E$ is the energy input to the system, $m$ is the mass, $g$ is the standard gravity, and $d$ is the moved distance. Lower CoT indicates lower energy consumption. The energy input of the system mainly comes from the electrical power to supply the pneumatic pump (3V, 0.12A). We plot the relationship between the actuation frequency $f$ (double the frequency of the electrical power supply) and CoT for the soft bistable swimmers with different wingspan lengths as shown in **Fig. 6C**. For the studied frequency range (0.4 – 0.67 Hz), the CoT decreases with the increasing $f$ for different $S$. The fastest soft swimmer with an intermediate $S = 150$ mm shows the lowest CoT. However, as $f$ further increases to 1 Hz, its CoT also increases, showing a U-shaped CoT curve with $f$ (**Fig. 6C**). The lowest CoT of ~ 39 J/kg/m is achieved at $f$ ~ 0.67 Hz with a high speed of 3.4 BL/s.

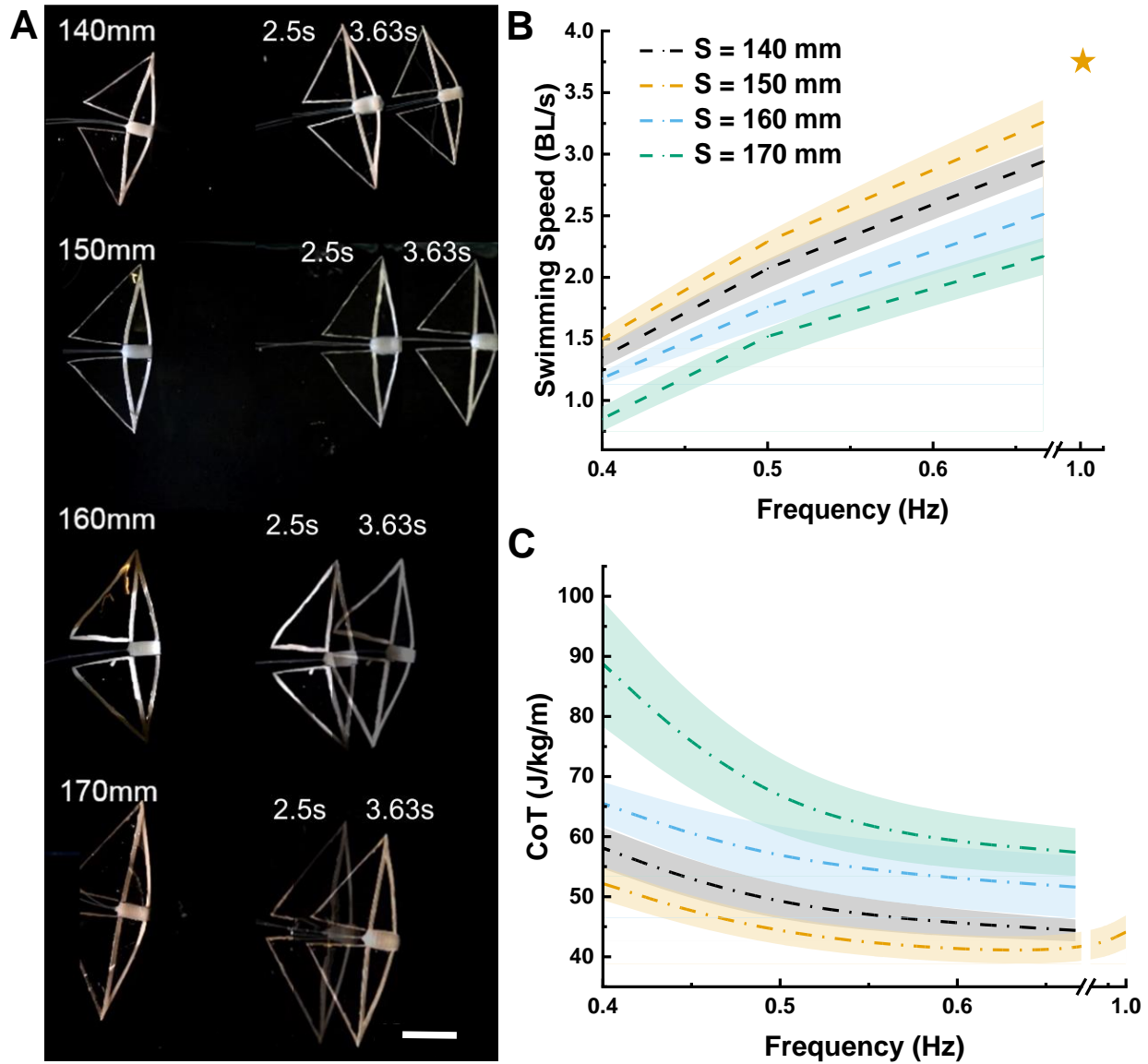

**Fig. 6. Bistable soft flapping-wing swimmers with both higher speed and lower energy-cost**. (A) Comparison of the swimming performance between the bistable swimmers with different *S* in top view of the overlapping time-lapse images. All actuators are pressurized at the same pressure of 55 kPa and actuated at the same frequency of 0.67 Hz. The swimmer with *S* = 150 mm shows the fastest speed (~ 3.4 BL/s, 0.67 Hz). The scale bar: 40 mm. (B- C) The relative speed and cost of transport (CoT) of the bistable soft swimmers with different *S* as a function of actuation frequencies. The orange star shows the fastest swimming speed (3.74 BL/s) for the swimmer with *S* = 150 mm under 1Hz actuation frequency. The swimmers with *S* = 150 mm shows both the fastest speed and the lowest CoT.

## Comparable fast and high-efficient swimming performances to biological counterparts

We further compare the swimming performance of our high-speed bistable soft flapping-wing swimmer (*S* = 150 mm) with the reported fast-swimming soft robots (*5-8, 10, 21-25*) and several speedy biological swimmers in wildlife(*36-43*) by categorizing them in a diagram of relative speed (BL/s) versus body mass in **Fig. 7A**. Only marine vertebrates are chosen for comparison due to the similar wave-like motions of their body, fin, and tail for propulsion (*44*). We note that the speed of the reported soft swimmers that are often monostable is entangled below ~ 1 BL/s (0.01- 0.69 BL/s) (*6-8, 10, 21-25*). Compared to the fastest speed (0.78 BL/s) of our reported bistable fish-like soft swimmer that harnesses swing body motion (*5*), the bistable soft flapping-wing swimmer here can achieve 4.8 times faster speed (3.74 BL/s) under much lower actuation pressure (55 kPa *vs.*160 kPa) with over 18 times lighter weight (a mass of 2.8 grams *vs.* 51 grams). Amplified by the bistability and butterfly-like swimming posture, the relative swimming speed of our soft swimmer (1.5 - 3.74 BL/s) is even faster or comparable to some of the marine vertebrates such as manta rays (~ 1.73 BL/s), dolphins (~ 3.28 BL/s), and humboldt penguins (~ 4.50 BL/s) (**Fig. 7A**).

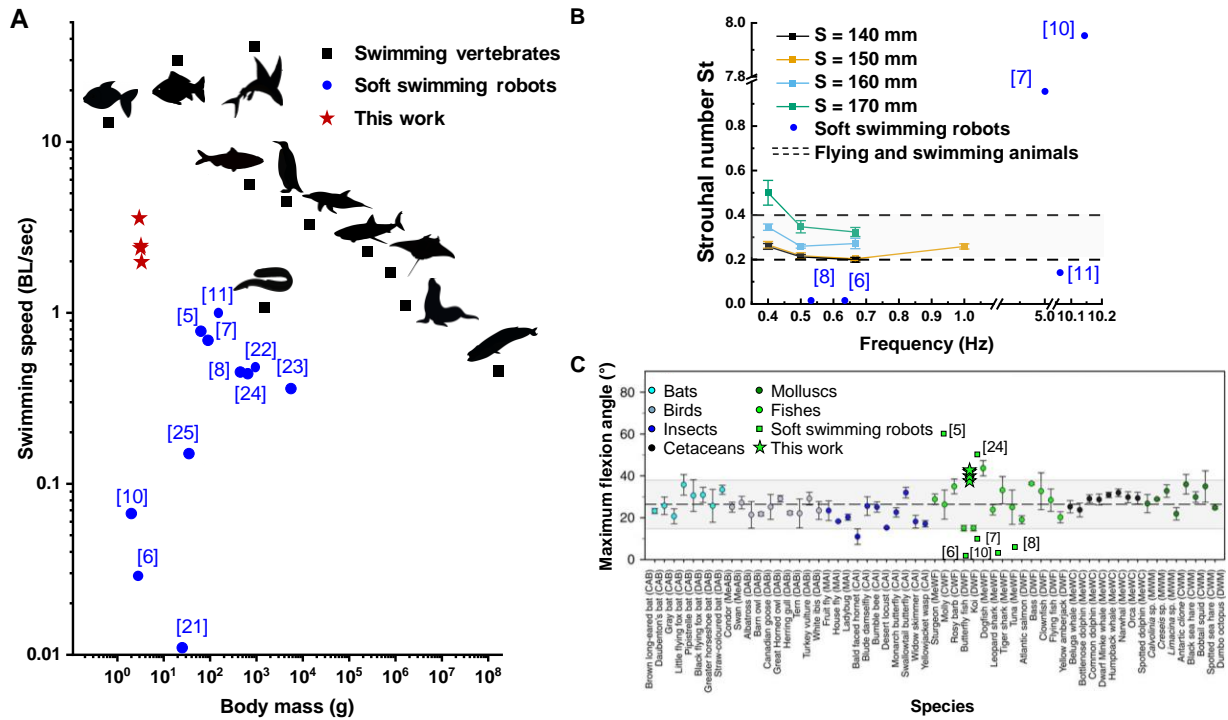

**Fig. 7. The high-speed and high-efficient bistable flapping-wing soft swimming robots ($S$ = 150 mm) with comparable performances to their biological counterparts.** (**A**) Comparison of the swimming speed (BL/s) vs. body mass between the proposed high-speed bistable flapping soft swimmers (red stars, mass of 2.8 grams, $S$ = 150 mm at frequencies from 0.4 to 1 Hz), some marine vertebrates, and the reported soft swimming robots in literatures (*5-8, 10, 21-25*). (**B**) Comparison of the Strouhal numbers St as a function of actuation frequency between the bistable flapping soft swimmers with different $S$ and the reported soft swimming robots in literatures (*6-8, 10, 11*). The two dashed lines define the lower and upper bound of the observed narrow range of the optimal St (0.2 – 0.4) in flyers and swimmers for maximum propulsion efficiency. (**C**) Comparison of the optimal range of maximum wing and fin flexion angles (~15° - ~ 40°) in flyer and swimmers (*45*) between the fast-speed bistable flapping soft swimmers ($S$ = 150 mm) and the reported soft swimming robots in literatures (*5-8, 10, 24*).

**Fig. 7B** further plots the propulsion efficiency of the fast-speed, bistable soft flapping-wing swimmers in terms of the Strouhal numbers St as a function of the actuation frequency $f$. We find that except the case of $S = 170$ mm actuated at 0.4 Hz, the Strouhal numbers St of all the studied bistable swimmers here fall within the observed optimal narrow range of $0.2 < St < 0.4$ (two parallel dash lines in **Fig. 7B**) in biological flyers and swimmers for the maximum propulsion efficiency (*3, 4*). It indicates that our soft swimmers can achieve both a high speed (3.4 – 3.67 BL/s) and a high propulsion efficiency ($0.2 \leq St \leq 0.255$). In contrast, the St numbers of the reported flapping or oscillating-based soft swimming robots in literature are either well above the upper bound or below the lower bound of the optimal naturally selected range, where the swimming speeds scatter from 0.02 BL/s to 0.78 BL/s actuated at a wide low-to-high frequency range of 0.5 – 10.13 Hz (*6-8, 10, 11*) (**Fig. 7B**). **Fig. 7C** shows another convergently optimal narrow range observed in a variety of flying and swimming animal propulsors in terms of the maximum wing and fin flexion angles (~15° - ~40°) (*45*), which are believed to contribute to the high efficiency of animal movements. Interestingly, we also find that the maximum flexion or flapping angle of our swimmers (~38°) also falls into its upper bound. In contrast, the reported soft swimming robots are either much higher or lower than the optimal maximum flexion angle range (*5-8, 10, 24*) (**Fig. 7C**).

**Multistable soft flapping-wing swimmer with enhanced maneuverability**

Despite the demonstrated high-performance soft flapping-wing swimmers that are comparable to their biological counterpart in terms of high speed and high efficiency, it can only achieve uni-directional forward swimming. To address the limitation, we further develop a maneuverable flapping-wing swimming robot that is capable of directional turning. As shown in **Fig. 8A** and **fig. S8**, similar to its bistable counterpart, the maneuverable swimmer is constructed from a multistable soft flapping actuator with two soft pneumatic bending actuators connected in parallel in the middle as the soft body (**fig. S2C-2D**), see Materials and Methods for details). The two bistable wings with wrapped thin films and extended flexible fins can be either independently flapped under single actuation for turning motion, or simultaneously flapped under dual actuations for forwarding motion with enhanced maneuverability.

**Fig. 8B** shows the overlapping time-lapse images of its navigation path marked by the dashed line (**Supplementary Video S9**). Its corresponding control sequence of the pneumatic flow rate is shown in **Fig. 8C**, where the pulsed amplitude and time for the flow rate are maintained at 1.2 liters/min and 0.172 s, respectively. Starting from a vertically placed posture in State I, upon solely actuated flapping of the left wing, the swimmer rotates clockwise from State I to State II, making a right turn, followed by an angled forward motion to State III under dual actuations of both wings. Then, it solely flaps its right wing to turn left to adjust its moving direction (State IV), followed by directional swimming from state IV to state V under dual actuation. We note that the multistable soft swimmer can generate a relatively large steering angle of ~ 25.5°/28.5° (clockwise/counterclockwise) per flapping stroke within 172 ms as shown in the inset of **Fig. 8D**, which corresponds to a fast tuning speed of ~ 157º/s. Such a speed is even higher than the recently reported fastest turning speed of 138.4º/s in the soft dielectric jumping robots on ground (*46*). In addition to the tunable swimming direction, similarly, its swimming speed can also be manipulated and accelerated by increasing the actuation frequency from 0.54 Hz (State IV to V) to 0.72 Hz (State V to VI) as shown in **Fig. 8B** and **Supplementary Video S9**. As expected, similarly, **Fig. 8D** shows that its swimming speed increases monotonically with the actuation frequency and can reach ~1.4 BL/s at 0.9 Hz. The compromised speed is due to its increased weight and modified design compared to its bistable counterpart.

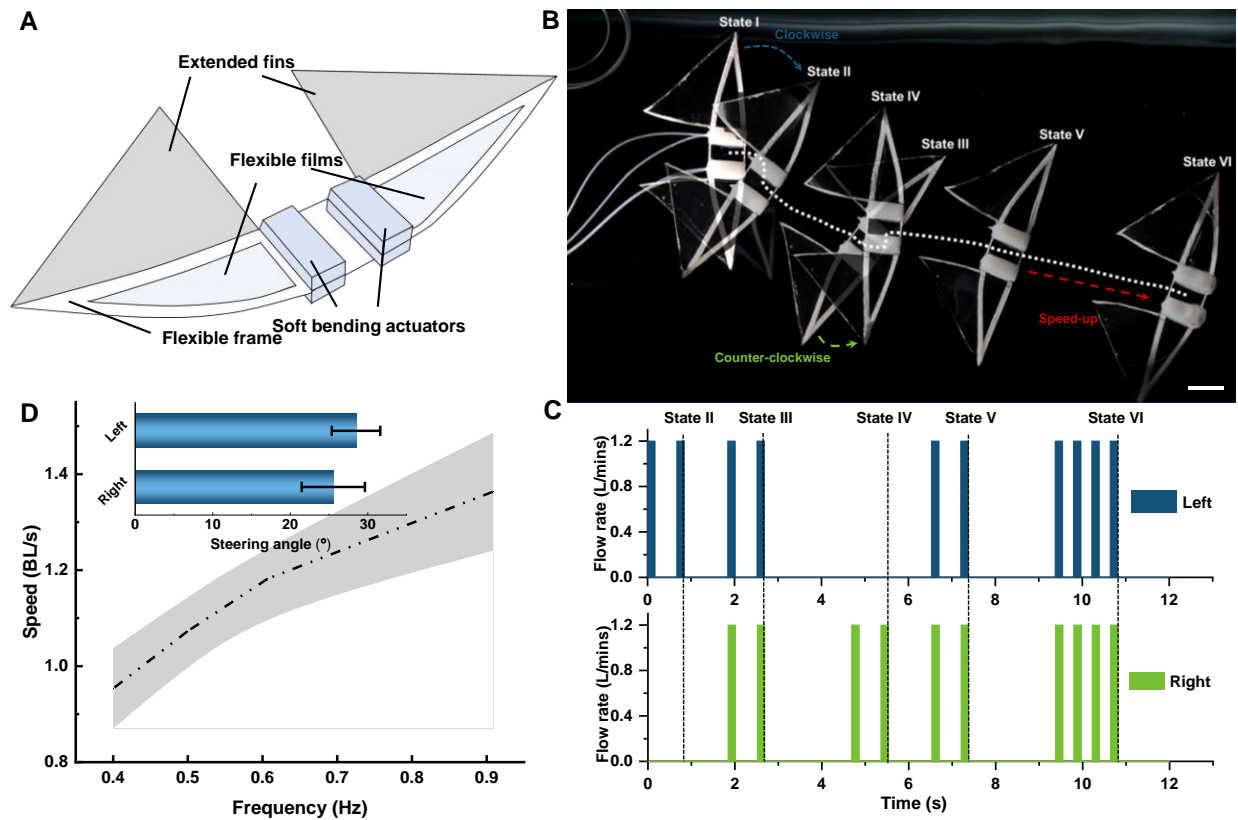

**Fig. 8. Multistable soft flapping-wing swimmer with enhanced maneuverability.** (**A**) Schematic of the multistable, maneuverable soft swimming robot by attaching the multistable flapping soft actuator with stretchable films and extended fins. (**B - C**) The overlapping time-lapse images of its navigation path through a right turn (State II), directional motion (State II - III), left turn (State IV), and speed-up directional motion (State IV – VI) in (**B**) with its actuation control of the flow rate in the left and right pneumatic bending actuator shown in (**C**). The scale bar: 20 mm. (**D**) The variation of relative speed of the multistable soft swimmer with the actuation frequency. The inset shows the steering angles under the single impulse of the flapping motion for the wings.

## DISCUSSION

In this work, we demonstrate harnessing snapping-induced amplified flapping and rotating of bistable and multistable flexible wings for achieving high-speed, high-efficient, and maneuverable swimming performance in a soft robotic swimmer. The achieved high swimming performances fall into the naturally selected optimal narrow design space for high propulsion and energy efficiency and are even comparable to that of their biological counterparts. The generic principle and simple flexible robotic structures presented in this work could be applied to other electric or stimuli-responsive actuations for small-scale transmission-free flapping-wing robots such as soft aerial robots or micro-air vehicles (MAVs), soft amphibious flying and swimming robots, and other jumping and kicking robots (**Fig. 9** and **Supplementary Video S10-S11**).

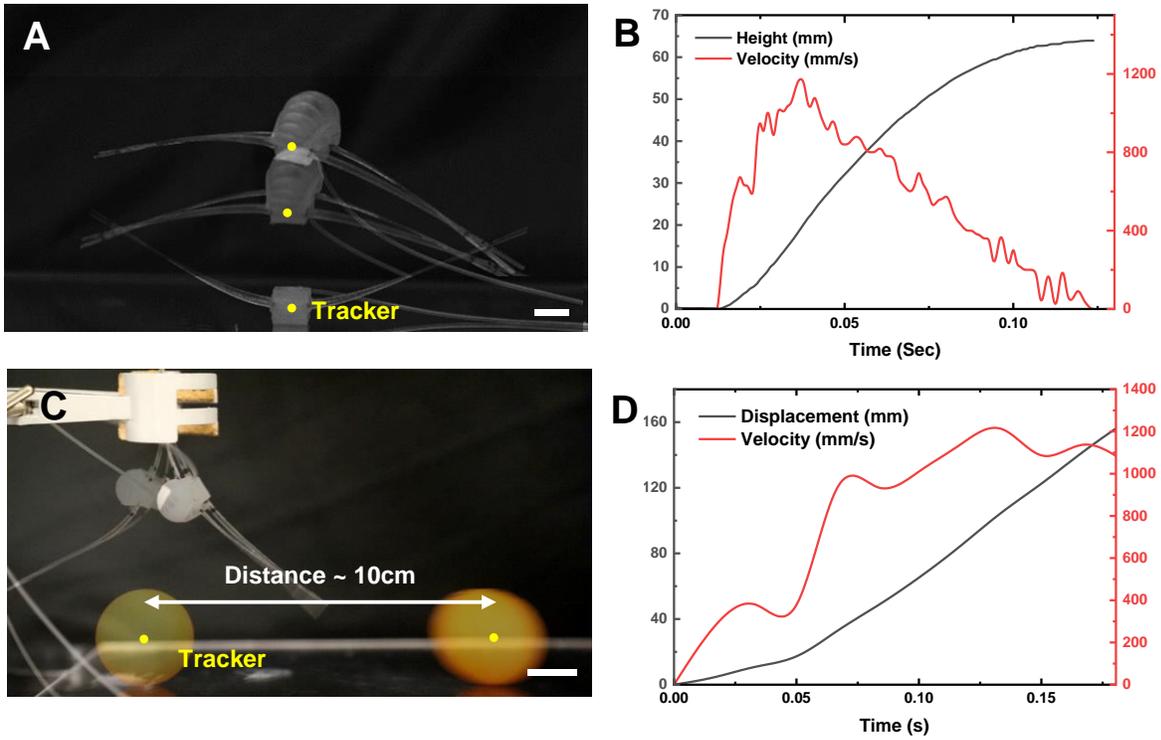

**Fig. 9. Other proof-of-concept applications of the bistable soft flapping actuators in soft jumper and kicker (a)** Jump-off process of the bistable flapping actuator from the ground under pressurizing the top pneumatic chamber. The yellow dot denotes the tracking point for motion. The scale bar: 4 mm. **(b)** Its jumping height and velocity profiles. It can achieve a maximum jumping height of 63 mm (~ 4.6 times its body height) in 0.12 s and a maximum jumping velocity of ~ 1,280 mm/s. It does not necessarily need a fast actuation speed rather a low air flow rate of ~ 20 ml/s in the soft body to drive the snapping of the wings. **(c)** Ping pong ball-kicking process of the bistable flapping actuator with one end fixed. The yellow dot denotes the tracking point for motion. The scale bar: 20 mm **(d)** The travelling distance and velocity profiles of the ball. It is capable of kicking a ping pong ball (2.7 grams) with a travelling distance of over 10 cm and a maximum travelling speed of over 1,200 mm/s.

Despite the promising results in this work, the full potential of the generic and scale-independent bistable and multistable flapping-wing mechanism for high performances still remains to be unleashed due to the intrinsic limitations of the soft pneumatic bending actuators such as low energy density and limited bandwidth. It also remains to be explored regarding applying its applications to other actuation mechanisms. These leave ample space for future studies and improvements.

First, achieving even faster swimming speed yet high efficiency at higher actuation frequencies is challenging for the current pneumatic bending actuator due to its limited bandwidth of about 1 Hz. Such a low bandwidth largely limits its applications to the scenarios that favor high frequency such as swimming and flying robots. In principle, for the observed $t_{snap}$ ~ 37 ms at $S$ = 140 mm, it could potentially allow an upper limit actuation frequency $f_{limit} = 1/(2t_{snap})$ ~ 13.5 Hz, thus, the full potential of snap-through instabilities has yet to be realized. The timescale of snapping in the bistable flapping-wing actuator $t_{snap}$ approximately scales with $S^2$ (*13*), i.e., $t_{snap} \propto S^2$. Thus, a relatively smaller size of $S$ for allowing higher actuation is favorable. For example, given the observed $t_{snap}$ ~ 37 ms at $S = 140$ mm, we expect that for insect-size flapping-wing flying robots with $S$ below 20 mm, $t_{snap}$ could be below 0.8 ms for allowing a potential high actuation frequency of up to 625 Hz. Thus, to achieve the high frequency, soft bending actuators with high bandwidth such as piezoelectric (e.g., PVDF bimorph structures over 100 Hz (*47*)) or dielectric actuators (over 10 Hz) will be preferred (*28, 29*) and explored in the future for potential applications in miniature soft flying robots and transmission-free flapping-wing MAVs (*48*). We note that the transmission-free flapping-wing design here could also largely reduce its self-weight by eliminating the complex compliant mechanism-based transmission systems often used in flapping-wing flying robots (*28, 29, 48*). In addition to the frequency, the critical force $F_{cr}$ from the soft bending actuator that drives the snap-through instabilities in the flexible flapping wing is important. We note that $F_{cr}$ approximately scales with $F_{cr} \propto E_r w_r d_{wing}(h_r/S)^3$ with $E_r$, $w_r$, and $h_r$ being the Young's modulus of wing ribbons, wing frame ribbon width, and ribbon thickness, respectively. For the studied flexible flapping-wing structure with $S = 140$ mm, $w_r = 3$ mm, $h_r = 0.55$ mm, we have $F_{cr}$ ~ 0. 8 N (fig. S5A). For insect-size robots with $S$ below 20 mm, to achieve a small critical driving force of below 0.1 N or even 10 mN that matches the scale of dielectric or piezoelectric actuators (*28, 29*), several means would be expected to take and explored in the future, including reducing $w_r$ and/or $h_r$ or using less stiff yet flexible materials.

Second, the demonstrated fast-speed swimming are still tethered to the air supply. Untethered actuation will be highly desired for achieving autonomous, high-speed soft swimming, jumping, and flying robots. The untethered systems could be achieved by either integrating miniaturized on-board sensing, controls, and power or utilizing remote actuation methods such as light or magnetic field (*49-51*).

Third, its body and wing shapes could be further optimized by exploring the complex flexible structure-fluid interactions through computer fluidic dynamics simulations and experiments to further increase the swimming speed and efficiency.

## MATERIALS AND METHODS
### Fabrication and actuation of bistable and multistable soft flapping actuators
The bidirectional pneumatic soft bending actuator was fabricated by following the typical manufacturing technique for fluid-driven soft actuators through molding and demolding approaches as shown in fig. S1. Ecoflex 00–50 (Smooth-on Inc) was used as the elastomeric materials for three sections of the soft bending actuator. The molds for two channel layers and intermediate layer were made of VeroWhite and 3D printed by Stratsys, Objet 260. Three cured Ecoflex layer were bonded by uncured Ecoflex 00-50. The flexible ribbons (ribbon width 3 mm) are made of polyester sheets (Grafix, thickness: 0.55 mm) and were laser cut to desired geometry and embedded into two edges of pneumatic actuator by Smooth-On SIL-Poxy Silicone Adhesive, two tips were bonded by Instant-Bond Adhesive 403 to form the bistable/ multistable structures. For experimental characterization of the geometries, dynamic performances, and swimming performances of the bistable soft flapping actuators and soft swimmers with different wingspan lengths ($S = 140$ mm, 150 mm, 160 mm, and 170 mm), the multistable flapping actuator and swimmer, and the monostable counterpart ($S = 150$ mm), triplicated prototypes were fabricated at minimum.

**Theoretical modeling of the buckled pre-curved wing shape**

The strain energy density U of the bistable flapping system can be expressed as
$$U = a\,\kappa_{YZ}{}^2 + b\,\kappa_{YZ}{}^2 + (1+c)\,\tau^2 \qquad (1)$$
by considering the bending energy in the two planes (the first two terms) and the torsion energy (the third term), where a and b are the rigidity ratios of the wing ribbon with $a = EI_1/GJ$ and $b = EI_2/GJ$. The strain energy in the soft body can be negligible compared to the wings considering its small bending deformation and much lower modulus. E and G denote the Young's modulus and shear modulus of the ribbon, respectively. $I_1$, $I_2$, and GJ are the principal moments of inertia of the cross-section of the ribbon and the torsional rigidity, respectively. c is the parameter denoting the energy stored in the actuator due to the twisting of the wing ribbon with $\tau$ being torsion. c is in the form of $c = GJL/E_a I_a S$, where $E_a$ and $I_a$ are the Young's modulus and the principal moments of inertia of the cross-section of the actuator. Based on the experimental measurements, the curvature and torsion of the wing's ribbon barely changes with the arc length. Thus, we use helical functions to approximate the shape of the wings and the curvature $\kappa$ and torsion $\tau$, which take the form of
$$\kappa = r/(r^2 + c^2),\ \tau = c/(r^2 + c^2) \qquad (2)$$
where the parameterization of the curve is expressed as $\tilde{r} = (r\cos t, r\sin t, ct)$. According to the Meusnier's theorem, $\kappa_1$ and $\kappa_2$ are the projection of $\kappa$ (44). The constraints are given by $\sqrt{r^2 + c^2}\,t = l$ and $r\cos t\cos\varphi + ct\sin\varphi - r\cos\varphi = w$, where $l$ and $w$ denote the length of the wing ribbon and the half length of the actuator, respectively. $\varphi$ is the rotation angle of the ribbon about the Y-axis and barely changes with the varying lengths of the ribbons. Note that we assume that the length of the actuator does not change when there is no air input. By minimizing the elastic energy stored in the system, its buckled wing shapes can be obtained.

**Dynamic block force measurement**

To quantify the flapping performance of the bistable actuator, we measured the dynamic block forces at the tips for different bistable flapping actuators with varied wingspan length $S$. The dynamic block force is defined as the impact force at the tips of bonded ribbons when its flapping motion actuated from one equilibrium state is blocked by a force sensor at another equilibrium state. The actuation pressure is set to 55 kPa, and the actuation frequency is set to 0.5 Hz for measurement. The flapping motion of the actuator generates a succession of impulses where the maximum stroke is determined as dynamic block force.

**Force-displacement curves characterization**

We characterized the static mechanical response of the bistable soft actuator with pre-curved wings through quasi-static indentation tests using Instron 5944 tensile tester. As schematically shown in the inset of **fig. S5**, one end of the pneumatic soft bending actuator, i.e., the soft body, was fixed and the other end, i.e., the head of the soft body, was vertically indented under a load control with a loading rate of 5 mm/min. The indentation forces and displacements were recorded and measured to plot the indentation force-displacement curves of the pre-curved soft actuators with different wingspan length S. To characterize the bistable bending behaviors of both wings, a pair of parallel indentation forces with a distance $d_m$ normalized by the projected wingspan length $l_w$ are applied to the two wings (**fig. S5**), the center of mass was fixed, the flexible ribbons was vertically indented under a load control with a loading rate of 5 mm/min.

The indentation forces and displacements were recorded and measured to plot the indentation force-displacement curves of the pre-curved soft actuators with different wingspan length *S*.

**Motion Capture**

The motions of studied bistable actuators and robots were captured by a high-speed camera (Photron SA-2) with the frame rate of 1000 fps. The motions are tracked by the customized markers on both tips of wingspan and analyzed through Photron FASTCAM Analyzer.

**Fabrication and actuation of soft swimming, jumping, and kicking robots**

For flapping-wing soft swimming robots, based on the fabricated bistable/multistable soft flapping actuators, stretchable thin film (3M Tegaderm transparent film roll, 16004) was wrapped around the bonded ribbons to form a membrane wing. On the trailing edges of both wings, triangular-shaped flexible films (width: 47.5mm, Scotch tape) were attached to the two flexible polyester ribbons as extended fins. For jumping soft robots, rather than bonding two end tips of the H-shaped flexible polyester ribbons, the two polyester ribbons were bonded at certain distance away from the end tips to form cross-shaped ends to increase the contact region with ground. A low air flow rate of ~20 ml/s was inflated into the top pneumatic chamber of the soft bending actuator with up-flapped wings to slowly bend the soft body (compared to snapping of two wings) and drive the passive snapping of the wings. For kicking soft robots, the fabricated bistable flapping soft actuator with wingspan length 140 mm was used as the ball-kicking soft robot. One end of the vertically placed bistable actuator was fixed to a rigid stand, the actuator was actuated under a pneumatic pressure of 55 kPa. The actuation timing control for the bistable and multistable actuators, swimmers, jumpers, and kickers is listed in Table S1. We follows our previous work on the experimental setup of the pneumatic inflation system for the soft robots and measurements of critical actuation pressure and flow rate for the bistable soft flapping actuators (*5*).

**Aquatic experimental setup**

The soft swimmers were place into an aquarium 121 × 32.3 × 31.6 cm and filled with 25 gallons of water. An open-looped pneumatic control system was tethered to the swimmer(*52, 53*). The swimming processes were filmed using Photron SA-2 (Stationary in-water test) and Canon 6D mark II. White colored LED was used as illumination.

**FEA simulation of bistable and multistable flapping actuators**

Parametric FEA simulation studies were conducted to investigate the formation of the pre-curved bistable and multistable actuators by bonding the tips with different wingspan length, as well as their actuated snapping-induced flapping performances under pneumatic actuations. The 3D geometric models of the H-shaped soft flapping actuators before bonding were built using the Solidworks software with the same measured geometrical dimensions as the prototypes. The commercial FEA software Abaqus was used for the FEA analysis with the Abaqus/Standard solver. The built geometric models of all the actuators in Solidworks were imported into Abaqus CAE as STL files and were meshed using the solid quadratic tetrahedral elements (C3D10). A mesh refinement study was conducted to verify the accuracy and convergence of the mesh. The flexible polyester ribbons are modeled with linear elastic materials with Young's modulus $E_{ribbon}$=1.54 GPa and Poisson's ratio of 0.44. The elastomer (Ecoflex-0050) is modeled as a hyper-elastic isotropic Yeoh model. The energy density is given by

$$U = \sum_{i=1}^{N} C_{i0}(\bar{I}_1 - 3)^i + \sum_{i=1}^{N} \frac{1}{D_i}(J-1)^{2i}$$

where $\bar{I}_1 = tr[dev(\mathbf{FF})^T]$, $J = \det(\mathbf{F})$, and $\mathbf{F}$ is the deformation gradient, $C_{i0}$ and $D_i$ are the material parameters. In our model, $N = 3$, $C_{10} = 0.019$, $C_{20} = 0.0009$, $C_{30} = -4.75 \times 10^{-6}$, $D_1 = D_2 = D_3 = 0$ (*52, 53*). Equal displacements ($d = 11.4$ mm) are applied in the opposite direction along the width of the pneumatic actuator to simulate the bonding process of the tips, which generated the buckled prestressed shapes without other constraints applied. The snap-through behavior is simulated using the dynamic implicit analysis with an applied pressure of 55 kPa. A damping coefficient (0.04) is added to prevent the vibration after reach to the other equilibrium state. Dynamic depressurization is performed after snapping. The minimum time step size is set to $10^{-12}$ to ensure the accuracy and well-capture of the snap-through induced flapping. The FEA simulation procedures of the multistable flapping actuator are the same as the bistable flapping actuator except the different initial geometric shape. The pressurization and depressurization of the four pneumatic channels were performed in sequence as shown in **Supplementary Video S2**.